\documentclass[
    ,final            
  ]
  {aipproc}

\layoutstyle{6x9}
\newcommand{\noun}[1]{\textsc{#1}}

\begin{document}

\title{Status of \noun{QGSJET}}


\author{Sergey Ostapchenko}{
  address={University of Karlsruhe, Institut f\"ur Experimentelle Kernphysik,
 76021 Karlsruhe,  Germany}
}

\begin{abstract}
Basic physics concepts of the \noun{QGSJET} model are discussed, starting from the general
picture of high energy hadronic interactions and addressing in some detail the
treatment of multiple scattering processes, contributions of ``soft'' and 
``semihard'' parton dynamics,  implementation of non-linear interaction effects.
  The predictions of the new model version (\noun{QGSJET~II-03}) are compared to selected accelerator data. 
Future developments are outlined and the expected input from the LHC
collider   for constraining  model predictions  is discussed.
\end{abstract}

\maketitle


\section{Introduction}
Nowadays hadronic Monte Carlo (MC) generators are actively used both in collider
and cosmic ray (CR) fields. The idea behind employing such MC models is twofold.
First of all, they provide a bridge between rigorous theoretical approaches and
corresponding experiments, thus allowing to confront novel ideas against data. 
On the other hand, MC simulation procedures are an inevitable
part of data analysis of modern experiments; an extraction of any new information
crucially depends on the understanding of the corresponding detector responce,
which is mimiced with the help of the MC tools.

Generally, MC models are developed on much less rigorous basis compared to the
underlying theoretical approaches and involve a number of additional, sometimes
{\it ad hoc}, assumptions in order to be able to treat complex phenomena studied
in experiments. 
On the other hand,  the ultimate goal
of  model development is to provide an adequate description of the
interaction features which are important for this or that particular study,
rather than to describe the interaction physics in its full complexity.

In the CR field, high energy MC generators are mainly used for the description
of the development of so-called  extensive air showers (EAS)  --
nuclear-electro-magnetic cascades induced by  primary cosmic ray (PCR) 
particles (to present knowledge, protons or nuclei) in the atmosphere.
 Studing various EAS characteristics,
primarily the position of the shower maximum $X_{\max}$ (the depth in g/cm$^2$,
 where maximal number of charged particles is observed) and the numbers of $e^{\pm}$
 ($N_e$) and of muons ($N_{\mu}$) at ground, one infers information on the PCR energy
 spectrum and  their composition. These techniques set a number of requirements
 on the corresponding MC generators. First of all, one is obliged to treat 
very general (minimum bias) hadronic collisions in the cascade. Furthermore, 
of special importance are model predictions for the inelastic hadron-nucleus
 interaction cross section, which makes a direct impact on the $X_{\max}$.
Going further, calculated EAS characteristics appear to be very dependent on the
model predictions for forward particle spectra, first of all, on the relative
energy difference between the initial hadron and the  most energetic secondary one --
the so-called inelasticity $K_{\rm inel}$. But the most crucial requirement to
CR interaction models is the predictive power, as one has to extrapolate the available
theoretical and experimental knowledge from the energies of present day colliders
up to the highest CR energies, which are as high as $10^{20}$ eV!

However, the problem under study offers also a number of simplifications.
For example, one is not  sensitive to ``fine'' details of hadronic 
final states, e.g., to production of short-lived resonances, whose effect is smoothed
out during the cascade development, or to the contribution of high $p_t$ hadrons,
which are typically collinear with the initial particle direction ($p_t/E\ll 1$)
and thus do not influence the lateral shape of the shower. The same applies to the
production of charm, bottom, or more exotic (e.g., SUSY) particles, for which
one obtains rather small inclusive cross sections (compared to pions and kaons)
to influence charged particle content of EAS, and which are produced mainly at
central rapidities, thus having a negligible influence on  $K_{\rm inel}$, hence,
on  $X_{\max}$. To resume, CR interaction models are generators
of very ``typical'' (mb level) interactions.

 \section{High energy interactions: qualitative picture}

The general picture for high energy hadronic collision is the one of a multiple
scattering process, being mediated by  parton (quark and gluon)
cascades proceeding between the two hadrons. 
 An inelastic interaction involves a 
number of elementary hadron production contributions,
where the coherence of the underlying parton cascades is broken and partons
fragment into secondary hadrons. In addition, it may contain
  multiple elastic re-scatterings, where
the coherence is preserved and all the partons from the corresponding chains
 recombine back to their parent hadrons, without a production of secondaries.
 In turn, elastic scattering is obtained when only
elastic sub-processes occur.

\begin{figure}[t]
\includegraphics*[width=7.2cm,height=2.3cm,angle=0]{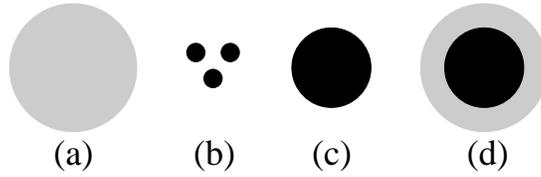}
\vspace{-.4cm}
\caption{Proton profile as viewed in soft (a), hard (b), semihard (c), and
 general (d) interactions.}
\label{fig:hadron-profile}
\vspace{-0.8cm}
\end{figure}

With the energy increasing, the number of elementary re-scattering processes
 grows rapidly, due to the larger phase space for parton
emissions. In addition, one expects a qualitative change in the structure of the
underlying parton cascades. At comparatively low energies, all the
 partons are characterized by  small transverse momenta; high 
 $p_t$ emissions are suppressed by the smallness of the corresponding running
coupling, $\alpha _s(p_t^2)$. By the uncertainty principle,   each 
emission  is characterized by  a large displacement of the produced parton in the transverse plane, $\Delta b^2\sim 1/p_t^2$.
 Thus, with the energy increasing further,
such ``soft'' parton cascades rapidly expand towards larger impact parameters,
 while the density of partons per unit transverse area remains small, the
hadron looking ``grey'', as depicted in Fig.~\ref{fig:hadron-profile}~(a).
 However, at sufficiently high energies an important contribution comes
from so-called  ``semi-hard'' and ``hard'' parton cascades, in which some
or all partons have comparatively high transverse momenta \cite{glr}. There,
the smallness of the strong coupling  $\alpha _s(p_t^2)$ is compensated
by a high parton density and by large logarithmic ratios of  the longitudinal
and transverse momenta for successive parton emissions. Purely hard cascades,
which start, e.g., from  valence quarks and contain only high $p_t$ partons,
do not expand  transversely,  $\Delta b^2\sim 1/p_t^2$ being small,
and lead to an increase of parton density in small areas  
in the transverse plane -- see Fig.~\ref{fig:hadron-profile}~(b), while giving
a negligible contribution to the total cross section. Contrary to that,
typical semihard re-scatterings are two-step processes: first, parton
 branchings proceed with a small momentum transfer and the cascade develops
towards larger impact parameters; next, high $p_t$ parton emissions become
effective, leading to a rapid rise of the parton density at a given
point in the transverse plane. As a result, the region of high parton density
extends to large impact parameters (Fig.~\ref{fig:hadron-profile}~(c)) and
the contribution dominates in the very high energy limit. General hadronic
interactions include all the mentioned mechanisms; hadrons in high energy
collisions look as shown in Fig.~\ref{fig:hadron-profile}~(d): there is an
extended ``black'' region of high density, dominated by the semihard processes,
and around it there is a ``grey'' region of low  density, formed by
purely soft parton cascading. In the ``black'' region one
expects strong non-linear parton effects to emerge, which result in the
saturation of parton densities and in the suppression of soft parton emissions
\cite{glr}.  On the other hand, such effects are weak
 in the ``dilute'' peripheral region.

\begin{figure}[t]
\includegraphics*[width=6cm,height=3.9cm,angle=0]{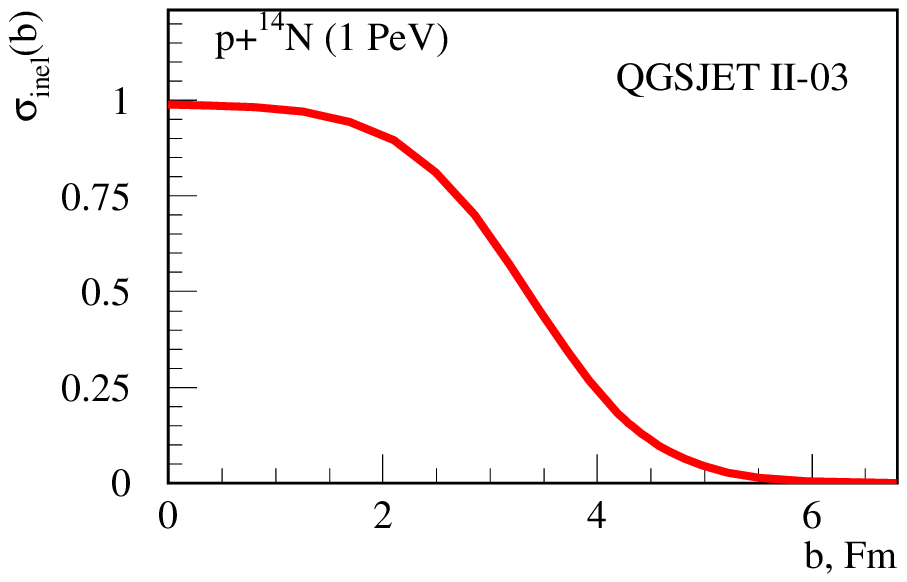} \hspace{1cm}
\includegraphics*[width=6cm,height=3.9cm,angle=0]{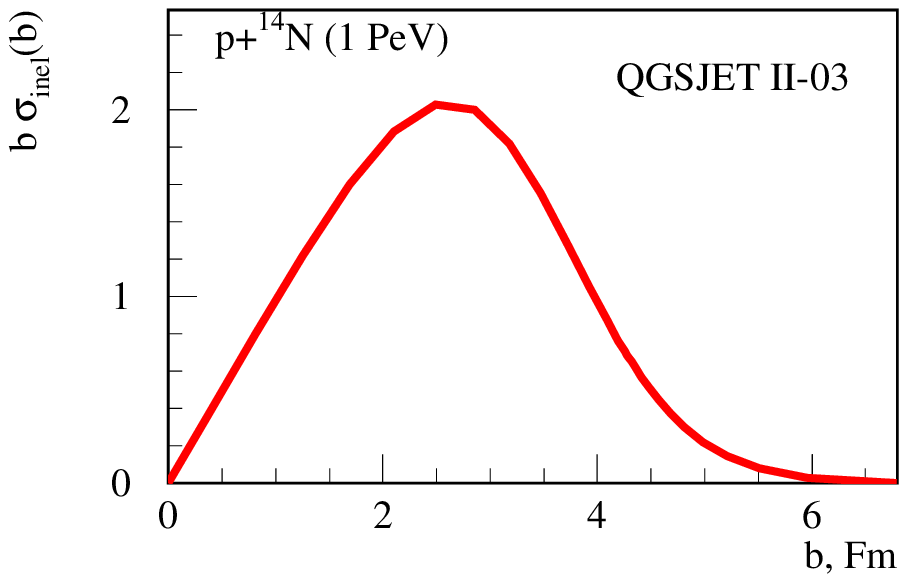}
\vspace{-.4cm}
\caption{Interaction profile   (left)  and relative 
contributions of various impact parameters to  the inelastic cross section 
(right) for proton-nitrogen collision at $10^6$ GeV.}
\label{sigprof}
\vspace{-0.8cm}
\end{figure}

What is the relative importance of the two regimes? 
As shown in Fig.~\ref{sigprof}~(left),
\begin{figure}[t]
\includegraphics*[width=6cm,height=3.9cm,angle=0]{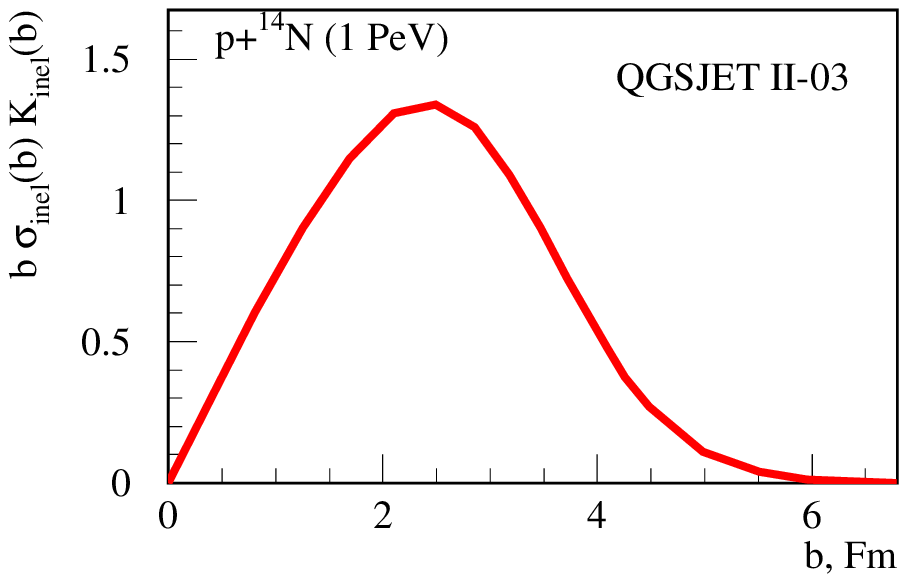}  \hspace{1cm}
\includegraphics*[width=6cm,height=3.9cm,angle=0]{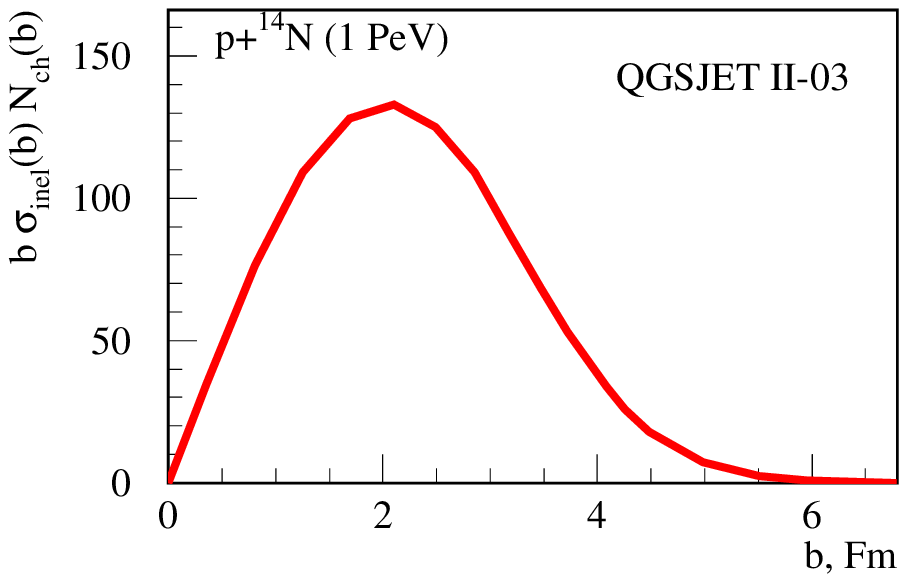}
\vspace{-.4cm}
\caption{Contributions of various impact parameters to $K^{\rm inel}_{pN}$
(left) and to $N^{\rm ch}_{pN}$  (right) at $10^6$ GeV.}
\label{kinprof}
\vspace{-0.8cm}
\end{figure}%
 in 1 PeV proton-nitrogen collisions
the ``black'' region (with the interaction
probability $\sigma_{\rm inel}(b)\simeq 1$) extends to impact parameters $b\sim 2$ Fm.
On the other hand, Fig.~\ref{sigprof}~(right) shows that most of the inelastic
collisions actually happen at  $b> 2$ Fm, i.e. they are mostly peripheral ones.
Similar conclusions follow if we investigate  relative contributions of 
central and peripheral interactions to the inelasticity $K^{\rm inel}_{pN}$
and the multiplicity of charged particles  $N^{\rm ch}_{pN}$ of the interaction,
as shown in Fig.~\ref{kinprof}:
 $K^{\rm inel}_{pN}$ is dominated by  peripheral collisions,
while for $N^{\rm ch}_{pN}$  central and peripheral contributions are
of equal importance.

\section{\noun{QGSJET} model}
General model strategy  to describe high energy  hadronic (nuclear) interactions
consists of the following steps: i) to describe an ``elementary'' interaction
(parton cascade), i.e.~to define the corresponding scattering amplitude and the
procedure to convert partons into final hadrons; ii) to treat multiple scattering
processes in the framework of the Gribov's Reggeon approach \cite{gri68};
iii) to describe particle production as a superposition of a number of  ``elementary''
 interactions. It is worth stressing that  the advantages of the pQCD formalism 
can be used here only partially and only at the first of the above-described steps,
as most of the elementary parton cascades develop partly (or even entirely) in
the nonperturbative ``soft'' region of small parton virtualities $|q^2|\simeq p_t^2$.

\begin{figure}[t]
\includegraphics*[width=7.cm,height=2.8cm,angle=0]{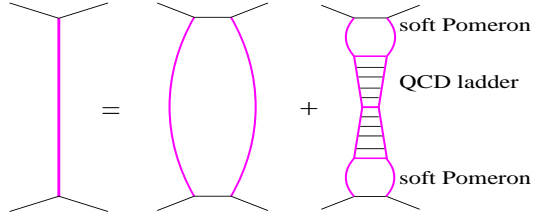}
\vspace{-4mm}
\caption{A ``general Pomeron'' is the sum of the ``soft'' and 
the ``semihard'' ones -- correspondingly the 1st and the 2nd graphs
in the r.h.s. \label{shp}}
\vspace{-0.8cm}
\end{figure}

In the particular case of the \noun{QGSJET} model \cite{kal94}, one introduces 
a cutoff $Q_0^2$ between the ``soft'' and ``hard'' parton dynamics, applies
DGLAP \cite{dglap} formalism for  $|q^2|\geq Q_0^2$, and employs a phenomenological
 ``soft'' Pomeron description for the nonperturbative ($|q^2|<Q_0^2$) parton
cascades. This way, an elementary interaction between hadrons $a$ and $d$
is described by a ``general Pomeron'' eikonal $\chi_{ad}^{\rm P}(s,b)$
(imaginary part of the corresponding amplitude), which consists of two terms:
the soft Pomeron eikonal $\chi_{ad}^{{\rm P}_{\rm soft}}(s,b)$,  representing
a pure nonperturbative (all $|q^2|<Q_0^2$) parton cascade,
 and the so-called ``semihard Pomeron'' eikonal $\chi_{ad}^{{\rm P}_{\rm sh}}(s,b)$:
$\chi_{ad}^{\rm P}(s,b)=\chi_{ad}^{{\rm P}_{\rm soft}}(s,b)+
\chi_{ad}^{{\rm P}_{\rm sh}}(s,b)$.
 The latter corresponds to a piece
of QCD parton ladder sandwiched between two soft Pomerons and represents a
cascade which at least partly develops in the region of high $|q^2|$, as shown in 
  Fig.~\ref{shp}  \cite{kal94} (see also \cite{dre99}).

To obtain  cross sections for various final states,
one makes use of the optical theorem, which relates the 
total sum of contributions of all final states to the imaginary part
of the elastic  amplitude for hadron-hadron scattering, hence, to the
contributions of various unitarity cuts of  elastic scattering diagrams.
The so-called Abramovskii-Gribov-Kancheli (AGK) cutting rules \cite{agk} 
 state that
only certain classes of cut diagrams are important in the high energy limit
and allow to relate such contributions to particular final states of interest.
For example, the non-diffractive cross section can be expressed as a sum
of contributions corresponding to a given number $n$ of elementary 
production processes (``cut''
Pomerons) and to any number of elastic re-scatterings (see  Fig.~\ref{multi-cut}):
\begin{figure}[t]
\includegraphics*[width=6cm,height=2.4cm,angle=0]{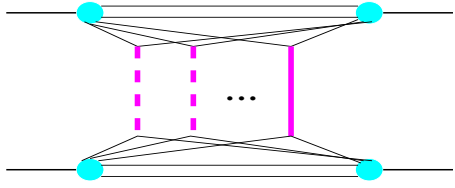}
\vspace{-4mm}
\caption{\label {mult} Typical inelastic interaction contains a number of
 elementary particle production processes, described by cut Pomerons 
 (broken thick lines), and any number of elastic re-scatterings --
uncut Pomeron exchanges (unbroken thick lines). \label{multi-cut}}
\vspace{-0.4cm}
\end{figure}%
\begin{eqnarray}
\sigma_{ad}^{\rm ND}\!(s)=\sum _{n=1}^{\infty}\sigma_{ad}^{(n)}\!(s)
=\sum _{n=1}^{\infty} \left\{\sum_{j,k}C_a^{(j)} C_d^{(k)}
\int\!\! d^{2}b  \;\frac{\left[ 2\lambda_a^{(j)}\lambda_d^{(k)}
\chi_{ad}^{\rm P}\!(s,b)\right]^n}{n!} 
\;  e^{-2\lambda_a^{(j)}\lambda_d^{(k)}
\chi_{ad}^{\rm P}\!(s,b)} \right \}, &&\nonumber \label{sigma-(n)}
\end{eqnarray}
where $C_a^{(j)}$ and $\lambda_a^{(j)}$ are correspondingly relative weights
and relative strengths of diffraction eigenstates for hadron $a$ \cite{kai79}.
 In turn,  elastic and diffractive
cross sections are obtained when elastic scattering diagrams are cut
between  Pomerons, with no one being cut,  selecting elastic 
or diffractive intermediate hadron states in the cut plane.

Finally, particle production is treated as a break out and a hadronization of strings
of the color field; in soft processes such strings are stretched between 
constituent  partons of the initial hadrons, to which
the Pomerons are coupled.
In case of semihard processes, one treats explicitely, within the DGLAP 
formalism, the production of partons (gluons and (anti-)quarks) of high 
transverse momenta ($p_t^2>Q_0^2$);
the strings are then formed between both these ``hard'' partons and 
the ``soft''  constituent  ones.

Importantly, the generalization from hadron-hadron
to hadron-nucleus and nucleus-nucleus interactions proceeds in a parameter-free
way, both for cross section calculations and for particle production  treatment
\cite{kal93}.

\section{Nonlinear effects (\noun{QGSJET II})}
Extending  model description to very high energies, one has to 
treat non-linear interaction effects, which come into play when individual
  parton cascades start to overlap in the corresponding phase space and 
to influence each other.
Such effects are expected to be extremely important at very high energies
and small impact parameters, i.e.~in the ``black'' region of high parton
 densities, where they lead to the parton density saturation \cite{glr}
and to significant reduction of secondary particle production. On the other hand, 
non-linear parton dynamics starts to manifest itself already at comparatively
low energies and large impact parameters, the experimental indication
being the rapid energy rise of the high mass diffraction cross section 
in the ISR energy range \cite{goul}.

\begin{figure}[t]
\includegraphics*[width=6cm,height=5cm,angle=0]{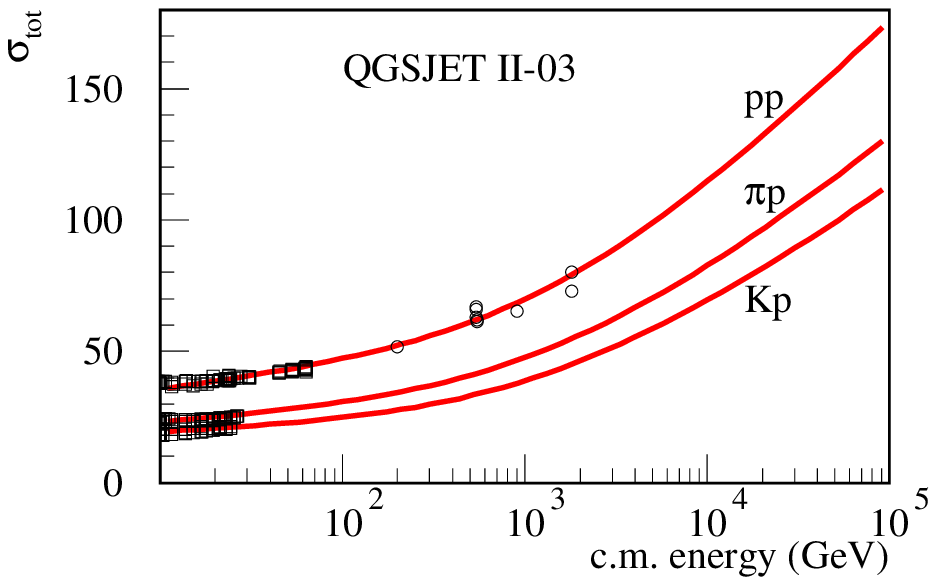} \hspace{.5cm} 
\includegraphics*[width=8cm,height=5cm,angle=0]{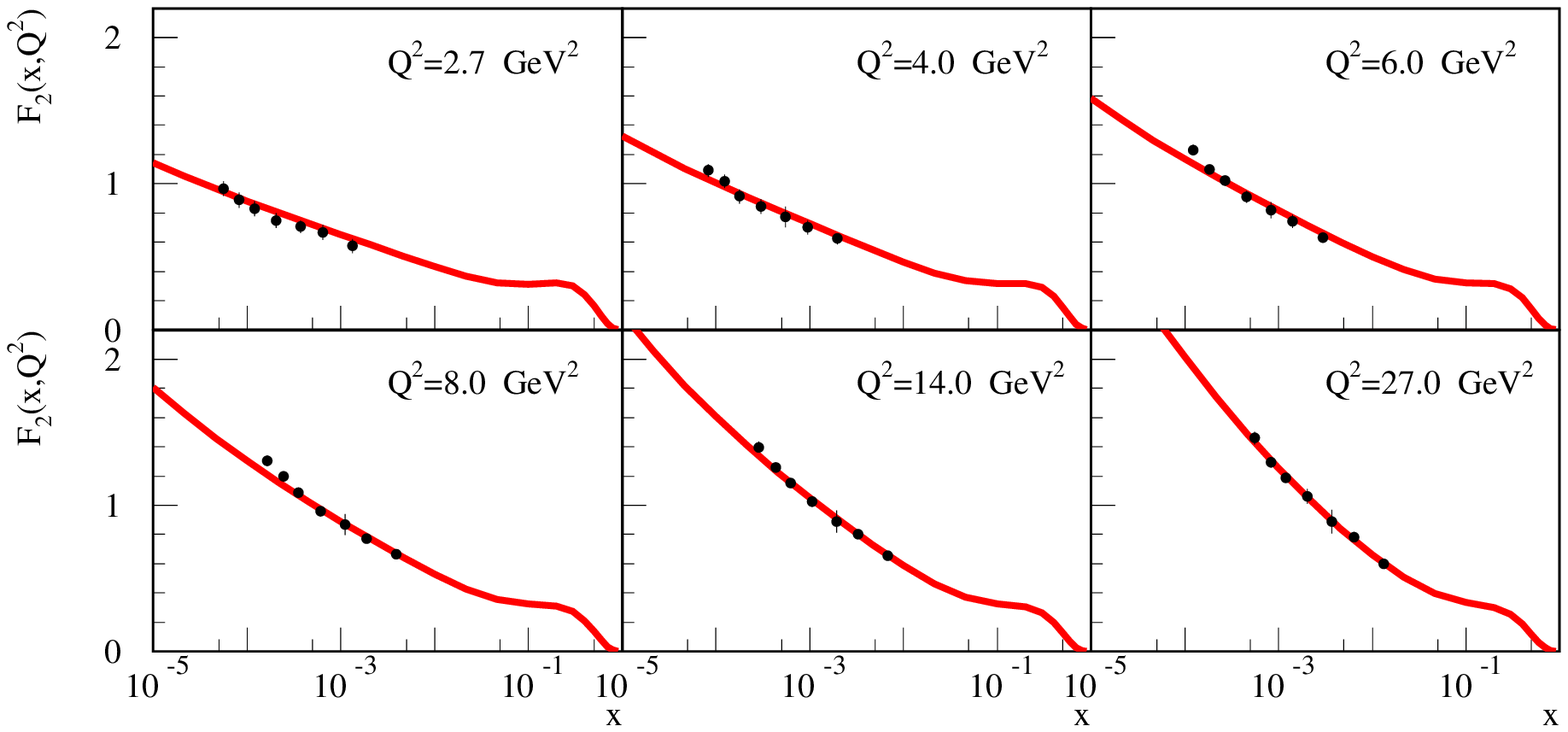}
\vspace{-.4cm}
\caption{Calculated total hadron-proton cross sections (left) and 
proton SF $F_2$ (right)
compared to experimental data (points) \cite{cas98,che05}.}
\label{f2}
\vspace{-0.8cm}
\end{figure}

Treating independent parton cascades effectively as Pomeron exchanges, the
corresponding non-linear effects are described  in the Reggeon Field Theory (RFT)
as Pomeron-Pomeron interactions \cite{kan73}. A resummation of the corresponding 
 (so-called enhanced) RFT diagrams has been worked out recently \cite{ost06},
 both for elastic scattering contributions
and for the corresponding unitarity cuts, and implemented in the QGSJET-II model
\cite{ost06c}. The basic assumption of the approach was that Pomeron-Pomeron
coupling proceeds  via parton processes at comparatively low
virtualities $|q^2|<Q^2_0$ and can be described using phenomenological multi-Pomeron vertices of eikonal type. The approach allowed one to obtain
a reasonable consistency with  experimental data on the total proton-proton
cross section and on the proton structure function  (SF) $F_2$ (see Fig.~\ref{f2}),
using a fixed energy-independent cutoff $Q_0^2=2.5$ GeV$^2$, i.e.~neglecting
possible parton saturation effects at higher virtuality  ($|q^2|>Q^2_0$) scales.

It is worth stressing that in the discussed approach, like in the usual linear 
scheme described in the preceeding Section, cross section calculations and particle production treatment are performed within the same framework.
 While the former are
expressed via hadron-hadron elastic scattering amplitude, which is obtained by
the resummation  of the underlying {\em uncut} RFT diagrams, the latter is based
on the computation of {\em unitarity cuts} of the very same diagrams, such that
summary contributions of cuts of certain topologies define relative weights
of particular final states. The corresponding results have been obtained in \cite{ost06} in the form of recursive equations, which allowed one to develop
 a MC algorithm to generate various  configurations of interactions,
including diffractive ones, in an iterative fashion.

It is noteworthy  that the obtained consistency between hadronic cross sections
and SFs,  demonstrated in Fig.~\ref{f2}, is highly nontrivial. By construction,
the discussed scheme  preserves the QCD factorization for inclusive jet spectra,
which allows one to describe correctly high $p_t$ hadron production.
However, the usual correspondence between the semihard  eikonal  
$\chi_{ad}^{{\rm P}_{\rm sh}}(s,b)$ and the inclusive cross section 
$\sigma_{ad}^{\rm jet}(s,Q_0^2)$ for the production of hadron jets of $p_t>Q_0$, 
namely $\int d^2b\; \chi_{ad}^{{\rm P}_{\rm sh}}(s,b)\equiv \sigma_{ad}^{\rm jet}(s,Q_0^2)$, which held in the original QGSJET model and which is still the 
cornerstone of all other MC generators,   is no longer valid in the nonlinear 
scheme. This is because there are two types of enhanced RFT diagrams which
contribute to  $\chi_{ad}^{{\rm P}_{\rm sh}}$. The diagrams of the first kind
provide factorizable corrections which can be absorbed in the parton distribution
functions (PDFs) and which do not violate the above relation. On the other hand,
the diagrams of the second kind are nonfactorizable ones and can not be accounted
for by any modification of the PDFs. As discussed in more detail in
 \cite{ost06,ost06c},
such contributions precisely cancel each other in the inclusive jet spectra,
the latter being expressed in the usual way via  universal PDFs of free hadrons
(which contain, however,  nonlinear corrections of the first kind).
However, there is no such cancellation in the interaction eikonal and correspondingly
in the partial cross sections for particular {\em exclusive} final states.
The latter are expressed via  nonuniversal (process-dependent)
PDFs, which are ``probed'' during interaction and are thus affected by the
surrounding medium   \cite{ost06c}.

\begin{figure}[t]
\includegraphics*[width=6cm,height=3.8cm,angle=0]{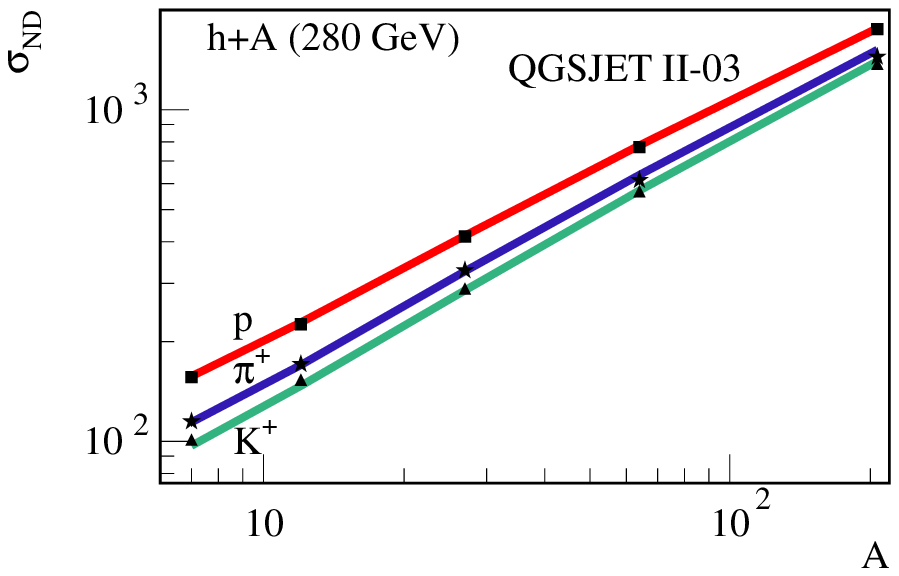} \hspace{1cm} 
\includegraphics*[width=6cm,height=3.8cm,angle=0]{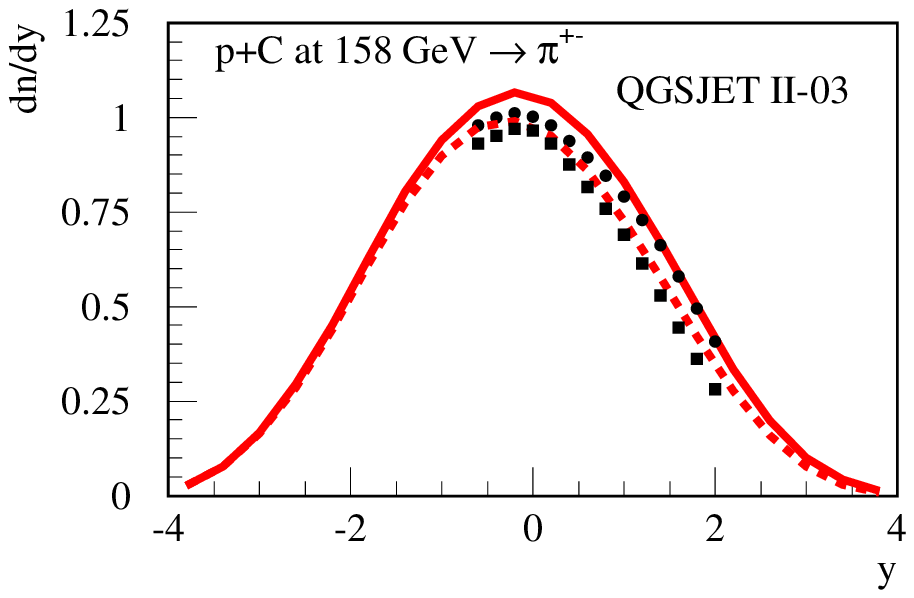}
\vspace{-.4cm}
\caption{Left: calculated $A$-dependence of  non-diffractive hadron-nucleus
cross sections at 280 GeV compared to experimental data (points) \cite{car79}.
Right:  calculated rapidity distributions of $\pi^{\pm}$ in $pp$ 
interactions at 158 GeV
compared to the data of the NA49 Collaboration \cite{na49}.}
\label{sig-nd}
\vspace{-0.8cm}
\end{figure}

Finally, the generalization from hadron-hadron
to hadron-nucleus (nucleus-nucleus) interactions is again performed
 in a parameter-free way. In the latter case, one obtains automatically
an $A$-enhancement of the screening effects, due to the coupling between
Pomerons connected to different nucleons. 
At fixed target energies, one obtains an impressive agreement with  measured non-diffractive hadron-nucleus cross sections (Fig.~\ref{sig-nd} (left));
 the predicted pion spectra also match the data reasonably well (Fig.~\ref{sig-nd} (right)).
\begin{figure}[t]
\includegraphics*[width=6.5cm,height=3.8cm,angle=0]{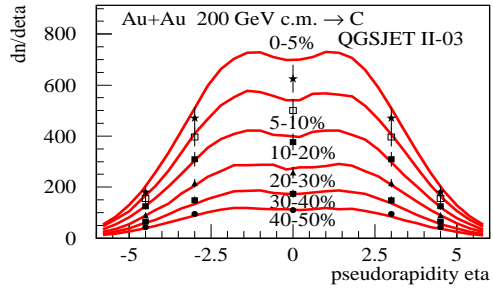}
\vspace{-8mm}
\caption{Pseudorapidity distributions of charged secondary particles
in Au-Au collisions of different ``centralities''; QGSJET-II
results (lines) are compared to  BRAHMS data \cite{brahms}. \label{rhicn}}
\vspace{-0.4cm}
\end{figure}

 Concerning  RHIC data  on central heavy ion collisions, those generally
can not be used
to judge on the goodness of a   MC model for CR applications --
because  the physics of very dense parton systems is of secondary importance for EAS
development. However, such comparisons help to understand the range of applicability
 of a particular model approach. In the \noun{QGSJET II} case, one may expect that the
model overestimates secondary particle multiplicity in such reactions -- as
additional screening corrections, connected to   Pomeron-Pomeron coupling
in the region of high parton virtualities ($|q^2|>Q^2_0$),
are not taken into consideration.   The obtained reasonably good agreement 
with the data   (Fig.~\ref{rhicn})
 indicates that such effects are yet small at the RHIC energies.

\section{Outlook}
The principal novelty of the \noun{QGSJET II} model is a microscopic treatment of 
nonlinear interaction effects in hadronic and nuclear collisions. In this aspect,
it provides a reference point for all other hadronic MC generators, where such
corrections are introduced at a much more phenomenological level, typically by
means of empiric energy-dependent parameterizations of some model parameters
(see, e.g.~\cite{eng99}).
Generally,   QGSJET-II  stays in a reasonable
agreement with  present accelerator data, including ones of the RHIC collider.
Moreover, the model predictions at much higher energies seem  to match
experimental EAS data as well \cite{haungs}.

Further model improvement is presently in progress to account for the contributions
of so-called Pomeron ``loop'' diagrams. Being
suppressed in the present scheme at small impact parameters \cite{ost06},
they may provide finite corrections to elastic scattering amplitude
at large $b$, thus influencing  model calibration on the basis of   observed 
$\sigma^{\rm tot}_{pp}(s)$.

From the experimental side, all present MC models will substantially benefit
from future measurements of  total and diffractive proton-proton cross sections
at the LHC by the CMS and FP420 experiments, the two quantities largely determining
the high energy extrapolation of model predictions. On the other hand, the
 forthcoming studies of forward neutron and photon production by the LHCf experiment
will allow to constrain model treatment of the hadronization process.

\end{document}